\documentclass{article} 

\usepackage[printonlyused,nolist]{acronym}
\usepackage{mwe} 
\usepackage{paralist}
\usepackage{hyperref}
\usepackage{authblk}
\usepackage{breakcites}
\usepackage{titlesec}
\titleformat{\subsection}[runin]
        {\normalfont\bfseries}
        {\thesubsection}
        {0.5em}
        {}
        [.]
\begin{acronym}
	\acro{CT}{Computed Tomography}
	\acro{CNN}{Convolutional Neural Network}
	\acro{IOU}{Intersection Over Union}
	\acro{DNN}{Deep Neural Network}
	\acro{RECIST}{Response Evaluation Criteria In Solid Tumors}
	\acro{NMS}{Non Max Suppression}
	\acro{DSC}{Dice Similarity Coefficient}
	\acro{FP}{False Positive}
	\acro{FN}{False Negative}
	\acro{TP}{True Positive}
	\acro{CPM}{Competition Performance Metric}
	\acro{SGD}{Stochastic Gradient Decent}
	\acro{ROI}{Region Of Interest}
	\acro{FROC}{Free Receiver Operating Characteristic}
	\acro{RPN}{Region Proposal Network}
	\acro{RCNN}{Region based Convolutional Neural Network}
	\acro{HD}{Hausdorff Distance}
	\acro{GT}{Ground Truth}
\end{acronym}

\title{Lung Nodules Detection and Segmentation Using 3D Mask-RCNN}
\begin{document}	
\author[1]{Evi Kopelowitz}
\author[1]{Guy Engelhard}
\affil[1]{Algotec LTD, a Carestream Co}
\affil[ ]{\textit { evik@algotec.co.il, guye@algotec.co.il}}
\maketitle

\begin{abstract}
Accurate assessment of Lung nodules is a time consuming and error prone ingredient of the radiologist interpretation work. Automating 3D volume detection and segmentation can improve workflow as well as patient care. Previous works have focused either on detecting lung nodules from a full \ac{CT} scan or on segmenting them from a small \ac{ROI}. We adapt the state of the art architecture for 2D object detection and segmentation, MaskRCNN, to handle 3D images and employ it to detect and segment lung nodules from \ac{CT} scans. We report on competitive results for the lung nodule detection on LUNA16 data set. The added value of our method is that in addition to lung nodule detection, our framework produces 3D segmentations of the detected nodules.
\end{abstract}


\section{Introduction}
Detection of lung nodules and accurate evaluation of their size are crucial for tracking cancer progression. Detecting the nodules is difficult since  nodules vary greatly in shape and texture, and non-nodules such as vessels, fibrosis, diffusive diseases etc. have similar appearance to nodules. Once detected, nodules size is currently evaluated using \ac{RECIST}. This measurement criteria relies on a linear measurement of the nodule along its largest axial slice. \ac{RECIST} is shown to be inferior to a volumetric measurement \cite{WBFBBC2012, HPOZMKG2016}. Nevertheless, it has become the standard of care since the time and effort required to manually delineate the 3D boundaries of nodules make such a workflow impractical for clinical applications.
Therefore, an automated system that detects nodules and segments their 3D volumes can improve patient care by providing better information on disease progression, as well as reducing the time taken by radiologists to assess a lung \ac{CT} study.

There is a large volume of work dedicated to detection of lung nodules on \ac{CT} scans using 2D and 3D architectures. See for example \cite{DBLP:journals/corr/DingLHW17, DBLP:journals/corr/abs-1811-08661, DBLP:journals/corr/SetioTBBBC0DFGG16}. Similarly, previous works used \ac{CNN}s to segment lung nodules from small \ac{ROI}s \cite{Nam2018LungNS,DBLP:journals/corr/FengYLA17,pmid30575046,DBLP:journals/corr/abs-1802-03584}.

\cite{DBLP:journals/corr/abs-1811-08661} combined segmentation signals with object detection tasks to improve detection rates in various implementations of 2D and 3D networks. However, to date, no one reported on both object detection and segmentation results derived from one, end to end, trainable network.
We propose to adapt the MaskRCNN model \cite{DBLP:journals/corr/HeGDG17}, which achieves state of the art results on various 2D detection and segmentation tasks, to detect and segment lung nodules on 3D \ac{CT} scans.
\section{Methods}
 \subsection*{Architecture}
MaskRCNN is a 2-stage object detector (\ac{RPN} followed by \ac{RCNN} and a semantic segmentation model (MASK)). We modify the 2D implementation of MaskRCNN \cite{matterport_maskrcnn_2017} to handle 3D images and to account for small object detection. Details regarding the full implementation of the model can be found in Appendix \ref{app:model} and \cite{mygit}.
 \subsection*{Training}
3DMaskRCNN is fully trainable end to end. Nonetheless, convergence is faster when training the backbone and \ac{RPN} together first, and then training only the second stage heads. Focal loss \cite{ DBLP:journals/corr/abs-1708-02002} and \ac{IOU} loss improve results in the class and MASK heads respectively. Training both segmentation and detection tasks simultaneously improves detection rate, similar to \cite{DBLP:journals/corr/abs-1811-08661}. We use dropout and heavy augmentation during training to avoid overfitting.
We perform 10-fold cross validation.
 \subsection*{Inference}
We scan each image with overlapping sliding windows. Overlapping boxes are filtered using \ac{NMS}. To reduce \ac{FP}s, we keep only boxes with a segmentation mask volume $> 0$. We use our in house lung mask CAD to remove nodules detected outside  of the lungs.
\section{Experiments and Results}
We tested our model on the LUNA16 challenge, taken from the LIDC/IDRI database \cite{LIDCref}, which includes 888 \ac{CT} scans. The reference standard of the challenge consists of all nodules $>=$ 3 mm accepted by at least 3 out of 4 radiologists \cite{pmid21452728}.

Detection evaluation is performed using the \ac{CPM}, defined as the average sensitivity at 7 predefined \ac{FP} rates: 1/8, 1/4, 1/2, 1, 2, 4, 8. Radiologists performance was evaluated in two cases: (1) Including only nodules $> 3$ mm and (2) including all nodules. Sensitivity and \ac{FP}s were  averaged over the 4 individual performances with respect to the other three. The results are summarized in table \ref{table:detec} together with state of the art method \cite{DBLP:journals/corr/DingLHW17} and ZNET, the winner of the LUNA16 challenge.

Note that our network achieves \ac{CPM} of 0.826 with a single inference step, beating the winning result of the challenge. Improved detection results (score of 0.86) were obtained by performing a second \ac{FP} reduction step, in which the model is fed with centered patches around proposed nodules.
Although \cite{DBLP:journals/corr/DingLHW17} reports on 15 Candidates per scan, 3DMaskRCNN achieves the highest sensitivity at 7-8  \ac{FP}s per scan since the average number of \ac{TP}s per scan $ < 2$.

\begin{table}[htbp]
  \caption{Comparison of Detection results}
  \begin{tabular}{llll}
  \bfseries Model & \bfseries Sensitivity & \bfseries \ac{FP}s/scan & \bfseries \ac{CPM}\\
  Radiologists ($>3mm$) &  0.75 &1 & NA\\
  Radiologists (all) & 0.85 & 5 & NA\\
  3DMaskRCNN (ours) \ac{FP} reduction &  0.936 &7& 0.86\\
  3DMaskRCNN (ours) &  0.932 & 8 & 0.826\\
  2DRcnn + 3DCNN \cite{DBLP:journals/corr/DingLHW17} &  0.946 & $<15$ & 0.891\\
  ZNET  \cite{DBLP:journals/corr/SetioTBBBC0DFGG16}&  NA &NA& 0.811
  \end{tabular}
  \label{table:detec}
\end{table}

Nodule segmentation results are shown in  Appendix \ref{app:figures}, Fig. \ref{fig:segmentation} demonstrating both small and large nodules as well as solid and ground-glass nodules.

Segmentation overlap is measured with the \ac{DSC}. Table \ref{table:seg} lists these results. Our results are comparable with radiologists' agreement (as calculated from their individual segmentations). Comparing to competing methods is difficult as other papers show segmentation results for predefined \ac{ROI}s whereas our results are over the full 3D scan. Note, that the test set used in \cite{Nam2018LungNS} contains only 113 nodules whereas ours has over 1000 nodules.

\begin{table}[htbp]
  \caption{Comparison of Segmentation results}
  \begin{tabular}{lll}
  \bfseries Model & \bfseries DSC\\
  3DMaskRCNN (ours) &  $70 \pm 10$ \\
  Radiologists &  $76 \pm 16$ \\
  CNN on diameter\cite{Nam2018LungNS} &  $79 \pm 19 $\\
  PN-SAMP-S1\cite{DBLP:journals/corr/abs-1802-03584} &  $74 \pm 3.57 $
  \end{tabular}
  \label{table:seg}
\end{table}

We evaluate the correlation between predicted segmentations volume and \ac{GT} and found a strong correlation of 0.96, indicating that volumes are indeed a reliable measurement for size.
The accuracy of the boundaries is assessed with \ac{HD} and is $2.49 mm\pm 2.05 mm$. The fact that the standard deviation is of the same size of the \ac{HD} suggests that this measurement may be irrelevant for small object segmentation.

\section{Conclusions}
We show that 3DMaskRCNN can achieve competitive results for both detection and segmentation tasks. We demonstrate strong correlation between predicted volumes and \ac{GT} and suggest that nodules volume evaluated and predicted by our model is a reliable measure of nodules size and may replace manual segmentation. As most of the \ac{FP}s our model detects seem like genuine nodules, see Fig. \ref{fig:fi_in_detection} in Appendix \ref{app:figures}, we believe that continued training/testing in order to further improve the \ac{CPM} will inadvertently cause overfitting to the LUNA dataset and hurt generalization on unseen studies as we are continually testing on the same test data. A known problem with the LUNA dataset is that the \ac{GT} is the intersection of 3-4 radiologists' detections, resulting in a very limited and strict dataset, very unlike a typical output of a single radiologist \cite{pmid21452728}. We plan on training 3DMaskRCNN on a wider dataset in order to generalize our results.
\section*{Acknowledgement}The authors thank Amir Yaacobi for his contribution in the implementation of the 3DMaskRCNN architecture as well as Ohad Silbert, Hadar Porat and Zahi Peleg for fruitful discussions. The authors acknowledge the National Cancer Institute and the Foundation for the National Institutes of Health, and their critical role in the creation of the free publicly available LIDC/IDRI Database used in this study.

\bibliographystyle{unsrt}
\bibliography{references}
\appendix
\section{3DMaskRCNN Model Architecture}\label{app:model}
The 3DMaskRCNN is composed of four parts: backbone, \ac{RPN}, \ac{RCNN} for classification and bounding box regression and another \ac{CNN} for pixel segmentation of objects, which we refer to as MASK.
\subsection*{Input}
Images are rescaled to a resolution of 0.5 mm per pixel, and cropped into patches of size $128^{3}$). To reduce \ac{FP}s, we concatenate positive and negative patches \cite{IAN}. Positive patches contain at least one nodule. The concatenated patches are normalized to have zero mean and unit variance.
\subsection{Backbone}
We implement the Inception Resnet v2 model \cite{DBLP:journals/corr/SzegedyIV16} in 3D.
In the reduction blocks we replace pooling layers with kernel dilation of 2 and 3 (known as Atrous kernels, \cite{DBLP:journals/corr/ChenPK0Y16}), thus achieving a wider field of view while maintaining full resolution. The output of the network is a feature map comprised of the output of the first and last reduction steps.
\subsection*{RPN}
The \ac{RPN} model operates on the feature map outputs of the backbone. Each pixel in the feature map is scanned with two anchors (of sizes 16 and 64 with a ratio of 1).
We apply a bounding box regression and classification on every anchor. \ac{RPN} proposals with scores greater than $0.1$ are passed to the \ac{ROI} Align layer. If no proposal scores higher than $0.1$ are found, the $10$ anchors with highest scores are passed. The low thershold was chosen to reduce \ac{FN}
proposals and anchors are considered positive(negative) if their \ac{IOU} with a ground truth box is greater(lower) than $0.5$($0.1$). The \ac{ROI} Align layer crops the proposals from the feature maps and rescales them to a fixed size.
\subsection*{RCNN and MASK}
\ac{RCNN} receives the aligned proposals and applies several convolution layers to predict final object classification and bounding box regression.
The MASK head of the network receives dilated \ac{ROI}s (5 mm on each edge) in order to get a wider view of the nodule. Then, several convolution and deconvolution layers are applied to predict pixel level nodule segmentation.

During training we add dropout layers in the \ac{RPN} and \ac{RCNN} class heads. Training parameters are:  $lr=0.01$ (reduced by half on plateau) for the backbone and \ac{RPN}. $lr=0.001$ (reduced by half on plateau) for the \ac{RCNN} and Mask heads. Momentum is set to 0.9 throughout training, and \ac{SGD} optimizer is used.
Each part is trained for 100 epochs. We perform 10-fold cross validation as required by the challenge in order to predict on the full data set. We implement our model using Keras Tensorflow. Please refer to \cite{matterport_maskrcnn_2017} to find the full detailed description of each layer in the Mask\ac{RCNN}.
\begin{figure}[htbp]
 \includegraphics[width=0.75\linewidth]{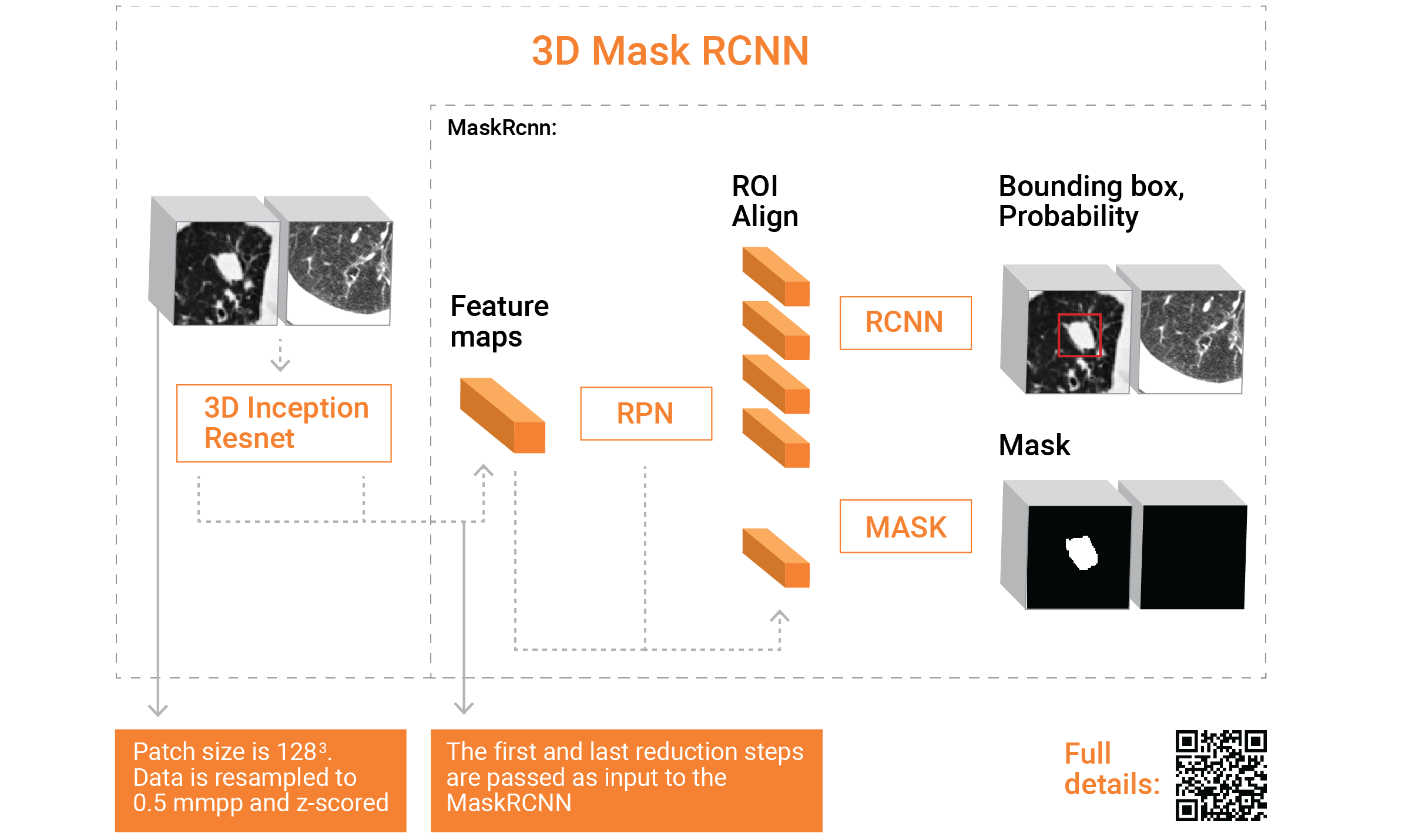}
 \caption{Diagram of the 3DMaskRCNN model}
 \label{fig:model}
\end{figure}

\section{Figures}\label{app:figures}
\begin{figure}[htbp]
 \includegraphics[width=0.75\linewidth]{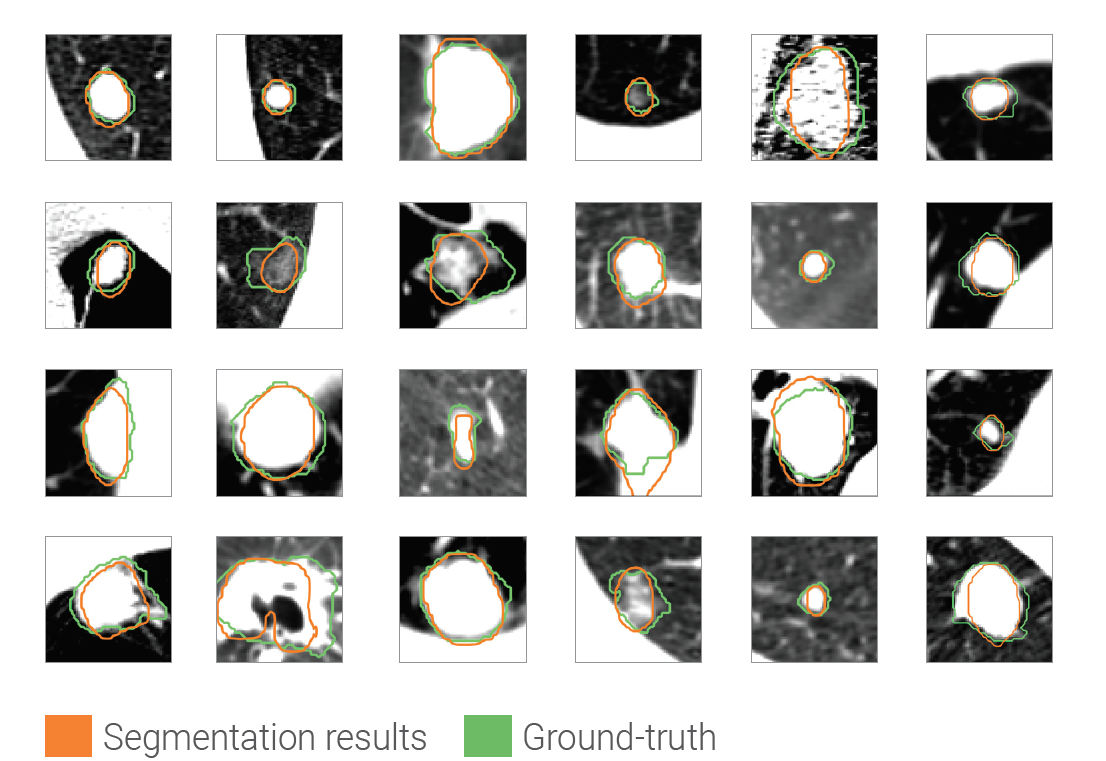}
 \caption{Examples of Nodule segmentation with 3DMaskRCNN. Box size is $3 cm^2$}
 \label{fig:segmentation}
\end{figure}

\begin{figure}[htbp]
 \includegraphics[width=0.75\linewidth]{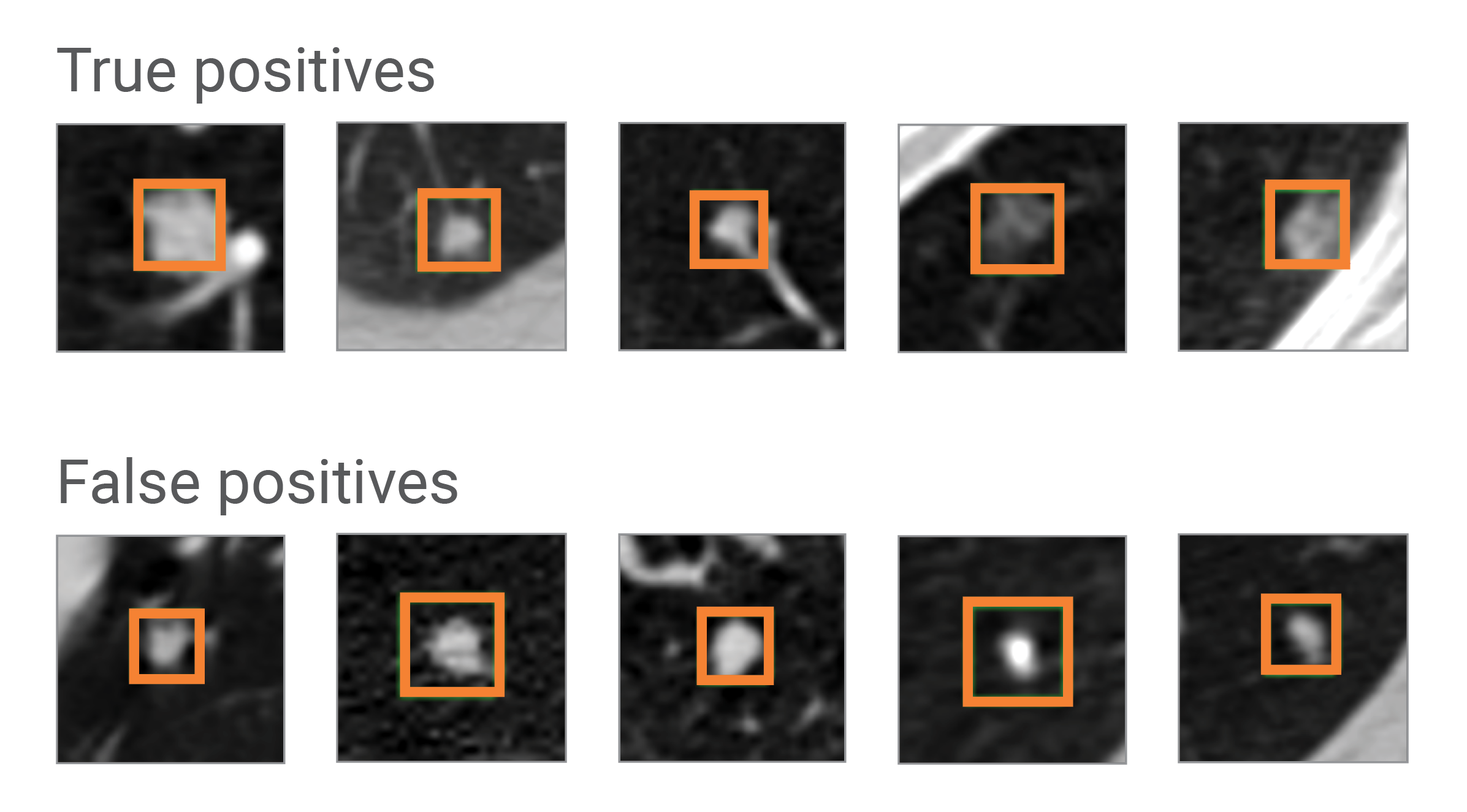}
 \caption{Examples of nodules detected by 3DMaskRCNN}
 \label{fig:fi_in_detection}
\end{figure}

\end{document}